\documentclass[12pt,preprint]{aastex}
 
\begin{document}

\newtheorem{theorem}{Theorem}[section]
\newtheorem{lemma}{Lemma}[section]
\newtheorem{corollary}{Corollary}[section]
\newtheorem{proposition}{Proposition}
\newtheorem{example}{Example}

\title{Magnetoelliptic Instabilities} 
\author{Norman R. Lebovitz}
\affil{Mathematics Department, University of Chicago}
\affil{Chicago, IL 60637, USA}
\email{norman@math.uchicago.edu}
\and
\author{Ellen Zweibel}
\affil{Astronomy Department and Center for Magnetic Self-Organization}
\affil{University of Wisconsin, Madison, WI 53706 USA} 
\email{zweibel@astro.wisc.edu}

\begin{abstract}
We consider the stability of a configuration consisting of a vertical
magnetic field in a planar flow on elliptical streamlines in ideal
hydromagnetics. In the absence of a magnetic field the elliptical flow
is universally unstable (the ``elliptical instability''). We find this
universal instability persists in the presence of magnetic fields of
arbitrary strength, although the growthrate decreases somewhat. We
also find further instabilities due to the presence of the magnetic
field. One of these, a destabilization of Alfven waves, requires the
magnetic parameter to exceed a certain critical value. A second,
involving a mixing of hydrodynamic and magnetic modes, occurs for all
magnetic-field strengths. These instabilities may be important in tidally
distorted or otherwise elliptical disks. A disk of finite thickness is stable
if the magnetic fieldstrength exceeds a critical value, similar to
the fieldstrength which suppresses the magnetorotational instability.

\end{abstract}


\section{Introduction}\label{intro}
The problem of momentum transport in accretion disks is widely
believed to require hydrodynamic or hydromagnetic turbulence for its
resolution. The origin of this turbulence may be sought in the
instability of laminar solutions of the equations of hydromagnetics,
solutions that are compatible with the geometry of accretion
disks. The recent history of these efforts has taken the form of first
recognizing such an instability mechanism, and then trying to
incorporate that mechanicsm into realistic disk models.

The magnetorotational instability (MRI) mechanism, originally discovered by
\citet{vel} and \citet{cha} and first applied to accretion disks in
\citet{bah}, is of this kind (see
\citet{mri} for a review). It appears in rotating, magnetized systems in
which the specific angular momentum increases outward and in which the
magnetic field is weak enough that rotational effects are not overwhelmed
by magnetic tension. 

A second mechanism, that of the elliptical instability considered by
\citet{jg} and others \citep{lpk,ryug, ryugv}, is also consistent with the
accretion-disk setting. This instability mechanism has been reviewed by \citet{rk02}. In the setting considered by Goodman et. al., it
appears to require a secondary in order to enforce departure from
rotational symmetry of the streamlines via a tidal potential. This is
certainly appropriate for binary systems but it is likely that, even
in the absence of a secondary, the laminar motion in the plane of the
disk would not be accurately circular, so the elliptical-instability
mechanism would appear to be a candidate of considerable
generality. It does not require a magnetic field. One of the
conclusions of the present paper is that it further persists in the
presence of a magnetic field. In the idealized setting of the present
problem, the latter may be of arbitrarily large strength. However, we
also argue that in the setting of a disk geometry, there may indeed be
a limit on the field strength.

In this paper we therefore investigate the interaction of a vertical
magnetic field with flow on elliptical streamlines, on the ground that
both magnetic fields and noncircular streamlines are likely
ingredients in accretion-disk settings. There are similarities with
and differences from previous work on effect of magnetic fields on the
elliptical instability \citep{rk94}, which are
discussed in \S \ref{discussion}.

\section{Formulation}\label{form}
We consider flow on elliptical streamlines together with a magnetic
field and investigate linear stability theory. The underlying
equations are the Euler equations of fluid dynamics
\begin{equation}\label{euler}
{\mathbf u}_t + {\mathbf u} \cdot \nabla {\mathbf u} = - \nabla p +
\left(\mbox{curl}\,{\mathbf B} \right) \times {\mathbf B}
\end{equation} 
and the induction equation
\begin{equation}\label{induc}
{\mathbf B}_t + {\mathbf u} \cdot \nabla {\mathbf B} = {\mathbf B}
\cdot \nabla {\mathbf u}. 
\end{equation} 
We shall assume that
div$\,{\mathbf u} =0$ and div$\,{\mathbf B} =0$ and that the fluid is
unbounded. 

It is easy to check that the following steady fields represent a
solution of the preceding system:
\begin{equation}\label{unpert.sol}
{\mathbf U} = \Omega \left(- \frac{a_1}{a_2}x_2,
\frac{a_2}{a_1}x_1,0\right),\;{\mathbf B} = \left(0,0,B\right), \; P =
\frac{\Omega ^2}{2}\left(x_1^2 + x_2^2 \right).\end{equation} Here
$\Omega$ and $B$ are constants, and a constant may also be added to
the pressure term. More general exact solutions of the combined
fluid/magnetic equations exist in an unbounded domain \citep{addc88};
the case in hand is probably the simplest of these.

\subsection{The perturbed system}\label{pert.sys}
Let ${\mathbf u}, {\mathbf B}, p$ be replaced by ${\mathbf U}+{\mathbf
u}, {\mathbf B}+{\mathbf b}, P + p$ in equations (\ref{euler},
\ref{induc}) above, and linearize. The resulting perturbation
equations are
\begin{equation}\label{pert.euler}
{\mathbf u}_t + {\mathbf U} \cdot \nabla {\mathbf u} + {\mathbf u} \cdot
\nabla {\mathbf U} = - \nabla p + \left(\mbox{curl} {\mathbf b} \right)
\times {\mathbf B}\end{equation} and
\begin{equation}\label{pert.induc}
{\mathbf b}_t + {\mathbf U} \cdot \nabla {\mathbf b} = {\mathbf B}
\cdot \nabla {\mathbf u} + {\mathbf b} \cdot \nabla {\mathbf U}, \end{equation}
together with the conditions that ${\mathbf u}$ and ${\mathbf b}$ be
solenoidal. These equations allow {\em rotating-wave} solutions of the
form
\begin{equation}\label{r-w.ansatz}
{\mathbf u}={\mathbf v}\left(t\right) \exp i \left({\mathbf
k}\left(t\right),{\mathbf x}\right),\;{\mathbf b}={\mathbf
w}\left(t\right) \exp i \left({\mathbf k}\left(t\right),{\mathbf
x}\right),\; p = \phi \left(t\right) \exp i \left({\mathbf
k}\left(t\right),x\right)\end{equation} where the expression $\left({\mathbf
k},{\mathbf x}\right)$ denotes the inner product.
Because ${\mathbf u}$ and ${\mathbf b}$ are solenoidal, the conditions
\begin{equation}\label{solenoidal}
\left({\mathbf k}\left(t\right),{\mathbf v}\left(t\right)\right) =0,\;\left({\mathbf k}\left(t\right),{\mathbf w}\left(t\right)\right) =0\end{equation}
must be satisfied.

Write
\begin{equation}\label{U.matrix}
{\mathbf U} = A{\mathbf x}\;\mbox{where}\;A=\Omega
\left(\begin{array}{ccc}0&-E&0\\E^{-1}&0&0\\0&0&0
\end{array}, \;\right), E = a_1/a_2.\end{equation} Then on substituting the
rotating-wave expressions from equation (\ref{r-w.ansatz}) into the
perturbation equations one finds
\begin{eqnarray}
\dot{\mathbf k}&=& -A^t{\mathbf k},\label{keq}\\
\dot{\mathbf v}&=& -A{\mathbf v} -i{\mathbf k}\phi + iB\left({\mathbf k}\times {\mathbf w}\right)\times {\mathbf e}_3 \label{veq}\\
\dot{\mathbf w}&=&i\left(k_3 B\right){\mathbf v} + A{\mathbf w}.\label{weq}\end{eqnarray}
Equation (\ref{keq}) can be solved to give
\begin{equation}\label{ksol}
{\mathbf k} = \left(\kappa \cos \left(\Omega t - \chi\right),
E\kappa \sin \left(\Omega t - \chi\right), k_3\right),\end{equation}
where $\kappa, k_3 $ and $\chi$ are constants.
The pressure coefficient $\phi$ can be eliminated with the aid of the
solenoidal condition (\ref{solenoidal}). One finds
\begin{equation}\label{phieq}
-i\left(\phi + Bw_3\right) = 2k^{-2}\left(A^t{\mathbf k},{\mathbf
v}\right).\end{equation} The equation for ${\mathbf v}$ now takes the
form
\begin{equation}\label{new.veq}
\dot{\mathbf v}= C\left(t\right){\mathbf v} +i\left(k_3 B\right){\mathbf w},\end{equation}
where
\begin{equation}
C\left(t\right) =-2\left(\Omega /k^2\right)
\left(\begin{array}{ccc}-E^{-1}k_1k_2&Ek_1^2&0\\-E^{-1}k_2^2&Ek_1k_2&0\\-E^{-1}k_3k_2&Ek_1k_3&0\end{array}\right)
+ \Omega \left(\begin{array}{ccc}0&E&0\\-E^{-1}&0&0\\0&0&0\end{array}\right).\end{equation}
It's convenient to break the six-dimensional system consisting of
equations (\ref{new.veq}) and (\ref{weq}) into two, one of size four and the other of size two:
\begin{eqnarray*}
\dot{v_1} &=&- \left(2\Omega /k^2\right)k_1\left(E k_1v_2 -
E^{-1}k_2v_1\right) + \Omega E v_2 +im w_1,\\ 
\dot{v_2} &=&
-\left(2\Omega /k^2\right)k_2 \left(E k_1v_2 - E^{-1}k_2 v_1\right) 
-\Omega E^{-1}v_1 + im w_2,\\ 
\dot{w_1} &=& im v_1  -\Omega
Ew_2,\\ \dot{w_2} &=& im v_2 + \Omega E^{-1}w_1,\end{eqnarray*}
where $m = k_3 B.$ These four equations are self-contained and the remaining equations,
\begin{eqnarray}\label{v3w3}
\dot{v_3}&=&-\left(2\Omega /k^2\right)k_3\left(Ek_1v_2-E^{-1}k_2v_1\right) +im w_3,\\
\dot{w_3}&=&im v_3 
\end{eqnarray}
may be integrated once the expression
\begin{equation}\label{c1}
c_1 \equiv Ek_1v_2 - E^{-1}k_2v_1
\end{equation}
is found by solving the four-dimensional system above. Equations
(\ref{keq}), (\ref{veq}),(\ref{weq}) and (\ref{phieq}) imply that 
\[ \frac{d}{dt} \left({\mathbf k}, {\mathbf v}\right)= im
\left({\mathbf k}, {\mathbf w}\right) \;\mbox{and}\;\frac{d}{dt} \left({\mathbf k}, {\mathbf w}\right)= im
\left({\mathbf k}, {\mathbf v}\right).\]
Thus in solving this system we need to impose the conditions
that these inner products are zero initially; this will thereafter
maintain the incompressibility conditions (\ref{solenoidal}).

The incompressibility condition provides an alternative way of finding $v_3$ and $w_3$ once the
equations for $v_1,v_2,w_1,w_2$ have been solved, provided that 
$k_3 \ne 0.$ The only cases for which $k_3$ can vanish are those for which the combinations
$k_1 v_1 + k_2 v_2$ and $k_1 w_1 + k_2 w_2$ are also found to vanish on
solving the four-dimensional system above. It is not difficult to show
that there can be no instability associated with such a solution (see
in particular the equivalent system (\ref{c_eq}) below). Accordingly,
we henceforth consider only perturbations with vertical wave number
$k_3 \ne 0.$  

\subsection{Change of variables}\label{new-variables}
We change to new variables to facilitate subsequent
calculations\footnote{The origin of this change of variables is
related to the existence of the rotating-wave solutions.}.
\begin{eqnarray}\label{new_vars}
c_1&=&Ek_1v_2 - E^{-1}k_2v_1, \nonumber \\ c_2&=&k_1v_1 + k_2v_2
\left(=-k_3v_3\right), \nonumber \\ c_3&=&Ek_1w_2 -E^{-1}k_2w_1,
\nonumber \\ c_4&=&k_1w_1 +k_2w_2 \left(=-k_3w_3\right).
\end{eqnarray}
This is a time-dependent (periodic) change of variables since
${\mathbf k}$ is periodic in $t$.
\noindent The equations to be solved take the form 
\begin{equation}\label{c_eq}\dot{c}=D\left(t\right)c\end{equation} in these variables, with (we have put $\Omega = 1$ here to agree with earlier conventions)
\begin{equation}\label{D_matrix}
D\left(t\right) =
\left(\begin{array}{cccc}-2\left(E-E^{-1}\right)k^{-2}k_1k_2&-2&im&0\\2k^{-2}k_3^2&0&0&im\\im&0&0&0\\0&im&0&0\end{array}\right).
\end{equation}

\subsection{General Considerations}\label{general}

The coefficients of the matrix $D$ depend on the phase angle $\chi$
appearing in the expressions (\ref{ksol}) above for the wave vector
$k$. For purposes of studying stability, we may set $\chi =0$\footnote{On the
other hand, for purposes of solving the initial-value problem, which
involves integrating over initial wave vectors, we would need to retain
it.}. This is easily seen by making the substitution $t^\prime =
t-\chi$, which eliminates $\chi$ from the equation. For the
remainder of this work we take $\chi =0.$

\medskip

The system (\ref{c_eq}) presents a Floquet problem (cf. \citet{ys}): the stability of
the chosen steady solution ${\mathbf U}, {\mathbf B}$ depends on
whether there are solutions of this system that grow exponentially
with time. This is settled by finding the Floquet multiplier matrix
$M$. The latter is defined as follows. Let $\Phi \left(t\right)$ be
the fundamental matrix solution of equation (\ref{c_eq}) that reduces
to the identity at $t=0$. Then, since the periodicity of $D$ is
$2\pi$, $M= \Phi \left(2\pi \right).$ If any eigenvalue $\lambda$
of $M$ has modulus exceeding one, this implies that there is indeed an
exponentially growing solution.

It is familiar in conservative problems that 

\begin{proposition}\label{quartet}
Whenever $\lambda$ is an
eigenvalue of the Floquet matrix, so also are its inverse $\lambda
^{-1}$ and its complex conjugate $\overline{\lambda}$. 
\end{proposition}
The first statement of this proposition is a typically a consequence of
canonical Hamiltonian structure, the second a consequence of the
reality of the underlying problem. However, the system (\ref{c_eq}) is
not canonical, and the matrix appearing in it is not real. We can
nevertheless establish these familiar properties of the eigenvalues
directly from the system (\ref{c_eq}), as follows.  The time-reversal
invariance of the physical problem is reflected in the existence
of a reversing symmetry $R = \mbox{diag}\,\left(1,-1,-1,1\right)$ of
the matrix $D$ above: $RD\left(-t\right) =- D\left(t\right)R$,
implying that whenever $c\left(t\right)$ is a solution so also is
$Rc\left(-t\right)$. Since the solutions of the Floquet problem have
the structure $c(t)=p(t) \exp \left(\sigma t\right)$, there must also
be a solution $p(-t) \exp \left(-\sigma t\right).$ But the eigenvalues
of the Floquet multiplier matrix $M$ are the values $\lambda = \exp
\left(2\pi \sigma \right)$. This shows that if $\lambda$ is an
eigenvalue of $M$, so also is $\exp \left(-2 \pi \sigma \right) =
\lambda ^{-1}$.

Similarly, under matrix transformation $S = \mbox{diag}\,
\left(1,1,-1,-1\right)$, $D$ goes to its complex conjugate. This shows
that $M = S \overline{M} S^{-1}$, i.e., that $M$ and its conjugate
have the same eigenvalues.

Immediate consequences of Proposition \ref{quartet} are the following:
first, in the stable case, eigenvalues of $M$ lie on the unit circle; second,
if, as parameters change, an eigenvalue is at the onset of
instability, it must have multiplicity two (or higher). The latter
conclusion is because, in the complex $\lambda$-plane, the dangerous
eigenvalue $\lambda$ must leave the unit circle simultaneously with
$\overline{\lambda}^{-1}$, which lies along the same ray as $\lambda
$ and therefore coincides with it when they both lie on the unit
circle. Thus a necessary condition for the onset of linear instability is a
resonance where two Floquet multipliers coincide.

\subsection{Parameters}\label{parameters}
There are three dimensionless parameters that figure in this
problem. We call them $\epsilon, \mu,$ and $\eta$:
\begin{equation}\label{equ:definitions}
\epsilon = \frac{1}{2}\left(E - E^{-1}\right), \; \mu = k_3/k_0, \;
\mbox{and} \; \eta = k_0B.\end{equation} 
Thus $\epsilon$ represents
the departure of the streamlines of the unperturbed flow from axial
symmetry. In these equations $k_0 = \sqrt{\kappa ^2 + k_3^2}$ and
represents the length of the wave vector if $\epsilon = 0$. The
magnetic parameter $\eta$ depends not only on the strength of the
unperturbed magnetic field but also on the wavelength of the
perturbation. 

In the matrix $D$ above, the magnetic field enters through the
parameter $m=k_3 B$ (which is a
measure of the magnetic tension
force), and we shall continue to use this notation for
the present, on the understanding that $m=\mu \eta$.  It is also clear
that we can use $E$ rather than $\epsilon$ to measure the departure
from rotational symmetry, and we shall do this in some cases.

\section{Analysis}

The Floquet matrix is $$
M\left(\epsilon , \mu, \eta \right) = \Phi \left(2\pi,\epsilon , \mu
, \eta \right).$$
One could map out the stability and instability
regions in the $\epsilon \mu \eta$ parameter space numerically by
integrating the system (\ref{c_eq}) systematically for many values of
these parameters. We in fact do this for a selection of parameter
values in \S \ref{numerical} below.  However, in this and the
following section we present the outlines and results of an asymptotic
analysis based on regarding $\epsilon$ as a small parameter (details
are presented in the Appendices). This is more revealing than the
numerical reults on their own. It is also of considerable importance
in interpreting the numerical results, and is quite accurate even
for values of $\epsilon$ that are not very small (cf. Figure \ref{hydromode} below).
The calculation proceeds in two steps; finding $M$, and calculating its eigenvalues.

\subsection{The Floquet Matrix}\label{floquet}

 The asymptotic
analysis is facilitated by the circumstance that, if we put $\epsilon
=0$, the coefficient matrix $D_0$ (say) of equation (\ref{D_matrix})
becomes constant. We find the eigenvalues and eigenvectors of $D_0$ in Appendix
\ref{subsec:zop}. There are two complex conjugate pairs of modes. One pair
reduces to the ordinary hydrodynamic modes in the limit $\eta\rightarrow 0$. The
second pair are magnetic modes with zero frequency at $\eta = 0$. In the weak
field limit, the ratio of magnetic to kinetic energy is ${\mathcal{O}}(\eta^2/4)$ for the modified hydrodynamic modes and ${\mathcal{O}}(4/\eta^2)$ for the
magnetic modes. If $\eta\gg 1$, the kinetic and magnetic energies are near
equipartition for both types of mode.
 
We now turn to a brief description of the asymptotic (or
perturbation) procedure.  In what follows the parameter $\eta$ will
be held fixed so, to simplify the notation, we suppress the dependence
of the Floquet matrix on this parameter: $M = M\left(\epsilon ,\mu
\right) = \Phi \left(2 \pi , \epsilon, \mu \right)$. We shall need the
Taylor expansion
\begin{equation}\label{M_exp}
M\left(\epsilon , \mu \right) = M\left(0,\mu _0 \right) + M_\epsilon
\left(0,\mu _0 \right) \epsilon + M_\mu \left(0, \mu _0 \right) \left(\mu -
\mu _0 \right) + \cdots ,
\end{equation}
where the dots indicate higher-order terms in $\epsilon$ and $\mu -
\mu _0$.  The reason for allowing variations with $\mu$ as well as
variations with $\epsilon$ is that the region in the $\epsilon \mu $
plane where instability occurs is typically a wedge with apex at a
point $\left(\epsilon, \mu \right) = (0, \mu _0)$ and boundaries $\mu
= \mu_0 + \nu _{\pm}\epsilon$, where the slopes $\nu _+$ and $\nu _-$
are to be found. (cf. Figure \ref{hydromode} below). We will therefore
consider $\mu$ of the form
\begin{equation}\label{mu_expansion}
\mu = \mu _0 + \nu \epsilon + \cdots
\end{equation}
where $\nu$ may be regarded as a fixed parameter to be chosen
later. This will lead us to the widest of wedges in the $\epsilon \mu$
plane, of width $O\left(\epsilon\right)$, excluding other wedges of
width $O\left(\epsilon ^m\right)$ with $m \ge 2$. These higher-order
wedges typically occupy a tiny fraction of the parameter space
(see Figure \ref{ho} below). As a result of the representation
(\ref{mu_expansion}) we may write
\begin{equation}\label{M_exp_2}
M = M_0 + M_1 \epsilon + \cdots
\end{equation}
where
\begin{equation}\label{M0andM1}
M_0 = M\left(0, \mu _0 \right) \; \mbox{and}\; M_1 = M_\epsilon
\left(0,\mu _0\right) + \nu M_\mu \left(0 , \mu _0 \right).
\end{equation}

The matrix $M_0$ is given in Appendix \ref{subsec:zop}, together with an
expression for $M_{\epsilon}$. We effect a change of basis, which diagonalizes $M_0$
and simplifies the calculation of $M_{\epsilon}$, in Appendix \ref{distinct}. The elements
of $M_1$ are calculated in Appendix \ref{subsec:elements}.

\subsection{The characteristic polynomial}\label{char_poly}
Denote by 
\begin{equation}\label{charpol}
p\left(\lambda, \epsilon \right) = \left|M\left(\epsilon
\right) - \lambda I \right|\end{equation} 
the characteristic polynomial of the
Floquet multiplier matrix $M\left(\epsilon \right)$ and its roots by
$\Lambda _1, \Lambda _2, \cdots$. A necessary
condition for stability is that each root lie on the unit circle. To
explore the onset of instability for various parameter values, we
shall obtain the characteristic polynomial $p$ in the form
\begin{equation}\label{p_expansion}
p\left(\lambda , \epsilon \right) = p_0\left(\lambda \right) +
p_1\left(\lambda \right) \epsilon + p_2\left( \lambda \right) \epsilon
^2 + \cdots \end{equation} and exploit our knowledge of the roots of
$p_0$ to obtain the roots of $p\left(\lambda , \epsilon \right)$ in
the form of a Puiseux expansion in $\epsilon$ \citep{hil}).

The nature of this expansion depends on the
multiplicities of the roots $\left\{\lambda _k \right\}$ of $p_0$, the
characteristic polynomial of the unperturbed Floquet matrix
$M_0$. These roots are given by the expressions $\lambda _k = \exp
2\pi \sigma _k$ where the $\left\{\sigma _k \right\}$ are the
eigenvalues of the matrix $D_0$ given in Appendix \ref{subsec:zop} (equation \ref{evals})
; they are all distinct. However, it is possible for the multipliers $\left\{\lambda _k
\right\}$ to be repeated even when, as in the present case, the $\left\{\sigma _k \right\}$ are
distinct: if $\sigma _k - \sigma _l = i k$ for an integer $k \ne 0$,
then $\lambda _k = \lambda _l$. 

\medskip
A necessary condition for the onset of instability is that there be a
double (or higher) root of the characteristic equation, and we henceforth
restrict consideration to the case of double roots\footnote{Higher
order zeros are not ruled out in this problem, since there are three
independent parameters, but we do not pursue this here.}. For
definiteness, we suppose $\lambda _2 = \lambda _1$. Then the Puiseux
expansion takes the form
\begin{equation}\label{puiseux} \Lambda _1 \left(\epsilon \right) = \lambda _1 + \epsilon ^{1/2} 
\beta _{1/2} + \epsilon \beta _1 + \cdots .\end{equation}
Substituting into the characteristic equation (and taking into account
that $p_0$ and $p_0^\prime$ both vanish at $\lambda _1$) yields for
the coefficient $\beta _{1/2}$ of the leading-order correction the
equation
\begin{equation}\label{mu1/2} \beta _{1/2} ^2 = -2p_1\left(\lambda _1 \right)/p_0 ^{\prime \prime} 
\left(\lambda _1\right).\end{equation}
The two values of $\beta _{1/2}$ give the generic expressions for the change in a double eigenvalue, yielding a pair of roots branching
from the double root $\lambda _1$. However, if $p_1 \left(\lambda _1
\right) =0,$ this expression is inadequate and one must proceed to the
next term in order to determine the effect of the perturbation on the
stability. Under the present assumptions, it is indeed the case that $p_1$ vanishes at
$\lambda _1$, as we show in Appendix \ref{subsec:charpoly}\footnote{This
``nongeneric'' behavior can be traced to the circumstance that the perturbation expansion takes place at a codimension-two point, i.e., where two
relations must hold among the parameters.}.

We must suppose then that the expansion of $p\left(\lambda , \epsilon
\right)$ is carried out to second order in $\epsilon$:
\begin{equation}\label{p_exp_2}
p\left(\lambda , \epsilon \right) = p_0\left(\lambda \right) +
\epsilon p_1\left(\lambda\right) +
\epsilon ^2 p_2\left( \lambda \right) + \cdots
\end{equation}
Then in the Puiseux expansion above $\beta _{1/2}
= 0$ so $\Lambda _1 = \lambda _1 + \beta _1 \epsilon + \cdots,$ and $\beta _1$ is found by solving the quadratic equation
\begin{equation}\label{beta1_eq}
\frac{1}{2} p_0^{\prime \prime}\left(\lambda _1\right)\beta _1 ^2 +
p_1^\prime \left( \lambda _1\right) \beta _1 + p_2\left(\lambda _1 \right) =0.\end{equation}

\medskip
In the case at hand, the common value of $\lambda _1$ and $\lambda _2$
lies on the unit circle. In order for the perturbed values of the
Floquet multipliers to lie {\em off} the unit circle (and therefore imply
instability), it is easy to verify that it is necessary and sufficient that $\beta _1/ \lambda
_1$ have a nonvanishing real part. Thus if we define
\begin{equation}\label{ratio}
\alpha = \beta _1 /\lambda _1 ,
\end{equation}
we have the following criterion:
\begin{proposition}\label{stab_crit}
Either $\alpha $ is pure-imaginary and we infer stability
(to leading order in $\epsilon$), or $\mbox{Re}\, \alpha \ne 0$ and
we infer instability.\end{proposition}
The magnitude of the real part of $\alpha$ is also related to the growthrate of the instability. If we define an instability increment 
\begin{equation}\label{incr}
\Delta = \left| \Lambda \right| -1,\end{equation}
then $\Delta = \epsilon \mbox{Re}\, \alpha$ and the growthrate is equal to $\Delta /2 \pi$, to leading order in $\epsilon$.

The long calculations that lead to the coefficients appearing in
equation (\ref{beta1_eq}) are carried out in the Appendices. In the
notation employed there, equation (\ref{beta1_eq}) therefore takes
the form
\begin{equation}\label{beta1_eq_3}
\alpha ^2 - \left\{\tilde{J}_{11} + \tilde{J}_{22} +\frac{2\pi
i}{\mu}\nu \left[\omega _1 + \omega _2\right] \right\} \alpha + \left|
\begin{array}{cc} \tilde{J}_{11}+ 2\pi \nu \sigma _1/\mu
&\tilde{J}_{12}\\ \tilde{J}_{21}&\tilde{J}_{22} +2\pi \nu 
\sigma _2/\mu
\end{array} \right| = 0,
\end{equation}
In equation (\ref{beta1_eq_3}), $\alpha$
has the meaning of Proposition \ref{stab_crit} above, and the symbols
$\tilde{J}_{ij}$ are defined in Appendix \ref{distinct} (equation \ref{Jij}). 
There are obvious modifications of this formula if $\lambda _k =
\lambda _l$ instead of $\lambda _1 = \lambda _2$.

\subsection{The Resonant Cases}\label{kne2}
The resonant cases for $\epsilon = 0$ (circular streamlines) are those
parameter values $(\mu,\eta)$ such that $\omega _j - \omega _l =k$,
where $k$ is an integer. We'll find that these can be written in the
form $\mu = f\left(\eta\right)$ (e.g., equation (\ref{pure_hydro_res}
below). Since $\epsilon = 0$, the $\mu$ values in question are those
that were designated $\mu _0$ in equation (\ref{mu_expansion})
above. We no longer need the designation $\mu _0$ and, in the relations
below, use the symbol $\mu$ in its place.

If $k \ne \pm 2$ the matrix $\tilde{J}$ is
diagonal (see equation \ref {newJtilde} below) and equation
(\ref{beta1_eq_3}) has the roots $\alpha = J_{11} + 2\pi i \nu \omega
_1/\mu$ and $\alpha = J_{22}+ 2\pi i \nu \omega _2/\mu$ where $\sigma
_j = i \omega _j$ for $j=1,2,3,4).$ It is easy to check that these
diagonal entries are pure-imaginary (cf. equations \ref{diagonal_eq}
and \ref{Tinverse} below) and therefore, in accordance with
Proposition \ref{stab_crit}, there is no instability to leading order
in $\epsilon$. We therefore now consider the only cases $(k=\pm 2)$ that
can lead to instability to this order.

\medskip
Recall that the original parameters of the problem are $\epsilon$ --
representing the departure from axial symmetry of the undisturbed
streamlines -- $\mu = k_3/k_0$ -- representing the vertical wavenumber
--and $\eta = k_0B$. The auxiliary parameters $m=\mu \eta$ and $q=\mu
\sqrt{1 + \eta ^2}$ are introduced to simplify the notation. With the
frequencies taken in the order
\begin{equation}\label{four_freqs} \left( \omega _1, \omega _2, \omega _3, \omega _4 \right) = \left( \mu + q, -\mu -q, \mu -q, -\mu +q \right) \end{equation}
the replacement $\mu \to -\mu$ results in the same frequencies in the opposite order.  We may therefore assume without loss of generality that
$\mu > 0$.  Further scrutiny of the formula (\ref{four_freqs}) shows that we need only consider the following four, distinct, $k=2$ resonances.

\subsubsection{Case 1. $\omega _1 - \omega _2 = 2$}\label{resonance1}
The resonant modes are those that reduce, when $\eta =0$, to the
purely hydrodynamic modes. This case therefore represents the
modification, due to the presence of the vertical magnetic field, of
the universal elliptical instability.  In this case $\omega _1 = -
\omega _2 =\mu + q=1$, implying that
\begin{equation}\label{pure_hydro_res} \mu = \frac{1}{1+ \sqrt{1 + \eta ^2}}.\end{equation} 
This ratio changes from $1/2$ at $\eta =0$ to $0$ as $\eta \to \infty$.
Evaluating the integrals defining $\left(\tilde{J}_{ij}\right)$ (equation \ref{Jij} below) and
solving equation (\ref{beta1_eq_3}) above, we find
\begin{equation}\label{alpha_res1}
\alpha ^2 = \left[\frac{\pi}{2}\left(1+\mu \right)^2\right]^2 - \pi ^2
\left[\frac{2\nu }{\mu} - \mu \left(1+\mu
\right)\right]^2. \end{equation} This has a maximum instability
increment (when $\nu = \mu ^2 \left(1+\mu \right)/2$) given
(suppressing a factor of $\epsilon$) by
\[ \alpha _{max} = \frac{\pi}{2}\left(1+\mu \right)^2.\]
In the pure-hydrodynamic limit $\eta = 0$ we find $\mu = 1/2$ and
therefore $\alpha _{max} = 9\pi/8$. Since the growthrate is given by
$\alpha \epsilon /2\pi$, this gives a maximum growthrate of $9\epsilon
/16$, in agreement with previous results obtained by other methods \citep{wal,jg}. In the limit $\eta \rightarrow \infty$ this
maximum instability increment tends to the finite limit $\pi /2$,
about half its value in the pure-hydrodynamic limit. 

This instability has a {\em bandwidth} $\left(\nu _+ - \nu _-
\right)\epsilon$ which is, for given $\epsilon$ and $\eta$, the length
of the $\mu$-interval for which the unperturbed configuration is
unstable. It is determined by the values of $\nu$ that make the real
part of $\alpha$ vanish. These may be read off equation
(\ref{alpha_res1}) above:
\[ \nu _+ = \mu \left(1+\mu \right) \left(1+3\mu \right)/4,\; \nu _- = - \mu \left(1 -\mu ^2\right)/4.\]
In the limit $\eta \to 0$ (the pure-hydrodynamic case) these give $\nu
_+ = 15/32$ and $\nu _- = -3/32$. These values of $\nu _\pm$ can also
be inferred from Waleffe's treatment of the pure-hydrodynamic case. In
the limit of large magnetic parameter $\eta$, the width of the band
tends to zero.

\subsubsection{Case 2. $\omega _1 - \omega _3 = 2$}\label{resonance2} 
The resonant modes consist of a hydrodynamic mode and a purely
magnetic mode (one frequency would vanish if $\eta = 0$).
In this case $q=1$, implying that $ \omega _1= \mu +1$, $ \omega _ 3= \mu -1$ 
and
\begin{equation}\label{equ:reshm}  
\mu = \frac{1}{\sqrt{1+\eta ^2}}.
\end{equation}

Thus the ratio $\mu$ changes from $1$ at $\eta=0$ to $0$ as $\eta \to
\infty$.
This represents a 'mixed mode', i.e., the resonance is between a
purely hydrodynamic mode and a purely magnetic one. Evaluating the
integrals and solving the quadratic (\ref{beta1_eq_3}), we find

\begin{equation}\label{alpha_res2}
\alpha= i\pi \nu \left( \frac{2 \nu }{\mu} -1+\mu ^2   \right) \pm \sqrt{D}
\end{equation}
where
\[ D = -\pi ^2 \left[ \frac{2\nu }{\mu} - \frac{1}{2}\left(1-\mu ^4\right)\right]\left[\frac{2\nu }{\mu} - \frac{1}{2}\left(1-\mu ^2\right)\left(3\mu ^2 -1 \right)\right].\]
If $D<0$ then $\alpha$ is pure-imaginary and the unperturbed configuration is stable, so instability prevails if and only if $D > 0$.

Instability can indeed occur and has its maximal increment when $\nu = \mu ^3\left(1-\mu ^2 \right)/2$. This maximal increment is (except for a factor of $\epsilon$)
\[ \left( \mbox{Re}\, \alpha \right)_{max} = \frac{\pi}{2}\left(1-\mu
\right)^2 .\]
This ranges from $0$ when $\eta = 0$ to $\pi /2$ as $\eta \to
\infty$. 

The width of instability band can be calculated as in the preceding case by finding the values of $\mu$ for which $D=0$. These are
\[ \nu _+ = \frac{1}{4}\mu \left(1 - \mu ^4 \right) \; \mbox{and}\; \nu
_- = \frac{1}{4}\mu \left(1-\mu ^2 \right)\left(3\mu ^2 -1 \right).\]

\subsubsection{Case 3. $\omega _4 - \omega _3 =2$}\label{resonance3}
These are the purely magnetic modes, which play no role if $\eta = 0$.
 For this case we have $q - \mu =1$ implying, for fixed $\eta$ that
\begin{equation}\label{equ:resmm}
\mu = \frac{1}{\sqrt{1+\eta ^2} -1}.
\end{equation}

Since $\mu $ cannot exceed $1$, this resonance can only occur for
sufficiently large values of $\eta $, namely
\begin{equation}\label{eta_big}
\eta > \sqrt{3}.\end{equation}
For a given wavelength $k_0$ this would require a sufficiently large
magnetic field $B$. When this
condition is satisfied, we have $ \omega _4= 1 =- \omega _3 .$
The formula for the instability increment $\alpha$ becomes
\begin{equation}\label{alpha_res3}
\alpha ^2=\left[\frac{\pi }{2}\left(1-\mu \right)^2 \right]^2 - \pi ^2 \left[\frac{2\nu }{\mu} + \mu \left(1 -\mu \right) \right]^2.\end{equation}
The maximum instability increment for given $\epsilon$ is (again
suppressing a factor of $\epsilon$)
\[ \alpha _{max} = \frac{\pi }{2}\left(1-\mu \right)^2,\]
which occurs for $\nu = -\mu ^2 \left(1-\mu \right)/2.$
It vanishes in the limit $\eta =0$ and tends to $\pi/2$ as $\eta \to \infty$. The upper and lower edges of the band of instability are expressed by the formulas $\mu = \mu _0 + \nu _\pm \epsilon$ where $\nu _\pm$ may be determined from equation (\ref{alpha_res3}) by requiring $\alpha$ to vanish. One finds
\[ \nu _+ = \frac{1}{4}\mu \left(1-\mu \right) \left(1-3\mu \right) \; \mbox{and} \; \nu _- = -\frac{1}{4}\mu \left(1-\mu ^2\right).\]
The bandwidth is therefore
\[ \nu _+ - \nu _- = \frac{1}{2}\mu \left(1-\mu \right)^2.\]
This vanishes both in the limit as $\eta \to 0$ and in the limit as $\eta \to \infty$. The maximum bandwidth occurs when $\mu = 1/3$ or $\eta = \sqrt{15}$.

\subsubsection{Case 4. $\omega _1 - \omega _4 =2$}\label{resonance4}

For this $\mu =1$. It is clear from the expressions for the matrix $J$
that the latter vanishes in this case. This leads to pure-imaginary values of $\alpha$ and therefore there is no
instability associated with this resonance.

\section{Numerical results}\label{numerical}
In this section we present a selection of numerical results. These are
obtained by integrating the system (\ref{c_eq}) to obtain the
fundamental matrix solution $\Phi \left(t,\epsilon,\mu,\eta\right)$
and evaluating it at $t=2\pi$ to get the Floquet matrix
$M\left(\epsilon,\mu ,\eta,\right)$. For fixed $\eta$, the eigenvalue
of maximum modulus is found as a function of $E$ (rather than
$\epsilon$) and $\mu$, and the regions of the $E \mu$ plane where this
maximum modulus exceeds one are distinguished. We have carried this
out to $E = 1.6$ ($\epsilon = 0.4875)$ in the figures although there is no
limitation on the size of $E$, or of $\epsilon$, in this method.

\begin{figure}
\plotone{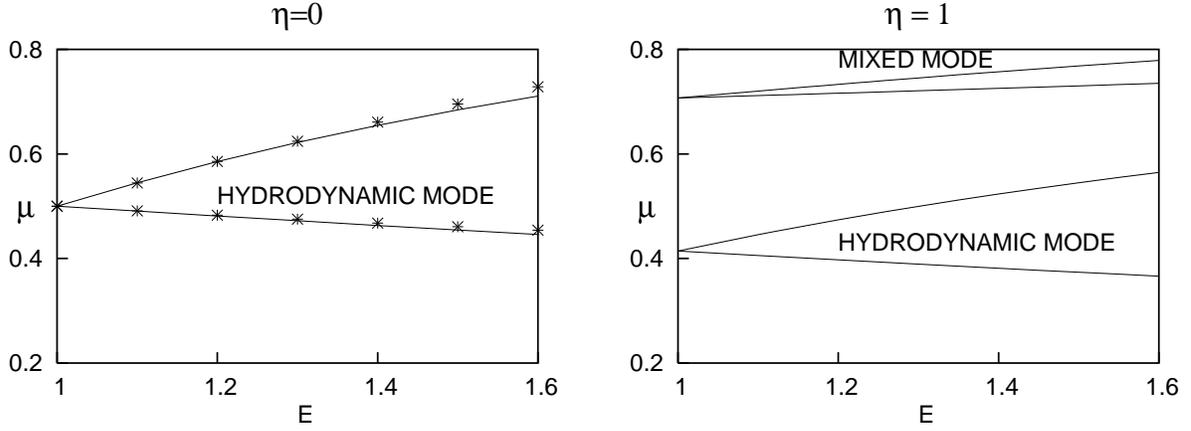}
\caption{\small On the left is the case $\eta =0$, i.e., the purely
hydrodynamic case considered earlier by several other authors. It is
shown here to contrast it with the case when the magnetic field is not
zero. The lines of asterisks indicate the same stability boundaries obtained from the asymptotic formulas $\mu = \mu_0 + \nu _{\pm}\epsilon$; this approximation is seen to be quite good. It is typical of the other cases considered. On the right is the case $\eta =1$, showing both the effect of
the magnetic field on the hydrodynamic mode and the existence of a new
mode of instability that is due to the presence of the magnetic
field.}\label{hydromode}
\end{figure}

\begin{figure}
\plotone{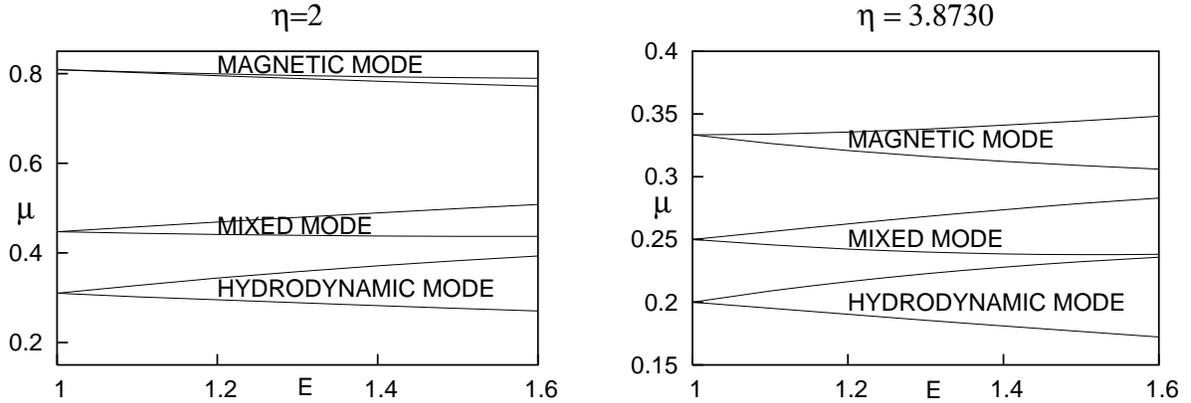}
\caption{\small On the left is the case $\eta =2$. Three regions of
instability occur here. The lowest of these is the modification of the
hydrodynamic-mode instability indicated in Figure \ref{hydromode}. The
middle region refers to the mixed hydrodynamic-magnetic mode. The
uppermost region, very thin and labeled MAGNETIC MODE, exists only for
values of $\eta$ exceeding $\sqrt{3}$, so is poorly developed for this
value of $\eta$.  On the right is the case $\eta = \sqrt{15} \approx
3.873$, for which the width of the uppermost, purely magnetic, mode
band is at its greatest. Note however that the vertical scale is
compressed relative to the diagram on the left, exaggerating the
widths of these bands by about a factor of two.}\label{mixedmode}
\end{figure}

In Figure \ref{hydromode}, we have taken the magnetic parameter equal
to zero in the left-hand panel, so this represents the purely
hydrodynamic case studied originally by \citet{bay} and \citet{rtp} and subsequently by others. For $\eta > 0$, a mixed mode of
interaction involving both hydrodynamic and hydromagnetic modes
comes into existence, and this is shown in the right-hand panel of
Figure \ref{hydromode}, where $\eta =1$. This is too small for the remaining
leading-order instability to appear.
\placefigure{hydromode}
That remaining instability, which represents a resonant interaction
between two modes that owe their existence to the presence of the
magnetic field, is indicated in Figure \ref{mixedmode} for a magnetic
parameter $\eta = 2$ (right-hand panel), slightly greater than the minimum value
($\sqrt{3}$) for the existence of this instability.
\placefigure{mixedmode}

\begin{figure}
\plotone{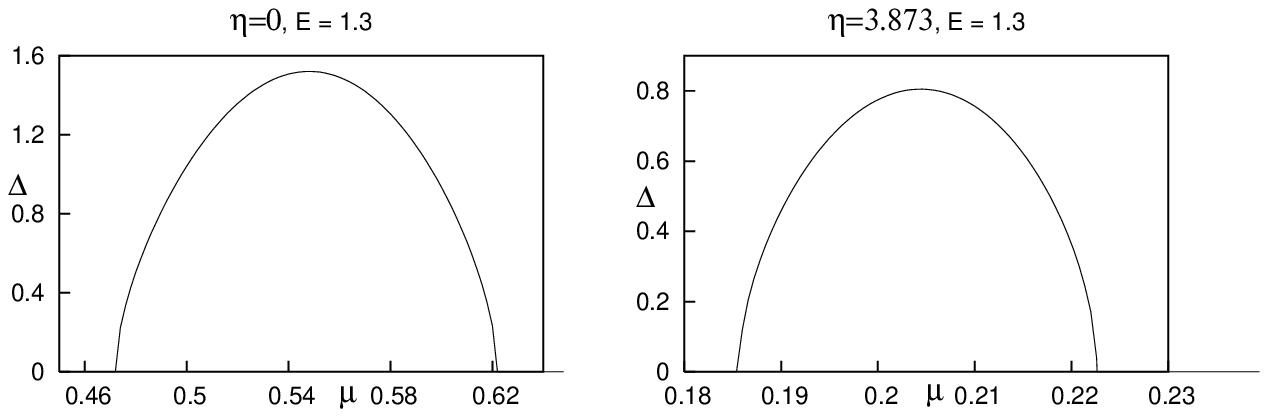}
\caption{The left-hand panel shows the instability increment $\Delta$ as a function
of $\mu$ for a fixed value of the magnetic parameter ($\eta =0$) and the ellipticity ($E = 1.3$). The right-hand panel does the same except that $\eta = \sqrt{15}$.}\label{grra}
\end{figure}

The asymptotic formulas imply that the maximal growthrate (or
equivalently the maximal instability increment $\Delta$) for each of
the wedges of instability tends to a fixed value as the magnetic
parameter $\eta$ increases. This is illustrated in Figure
\ref{grra}. However, the asymptotic formulas for the growthrates are
less accurate than those for the stability boundaries, for the larger
values of $E$. 
\placefigure{grra}

\begin{figure}
\plotone{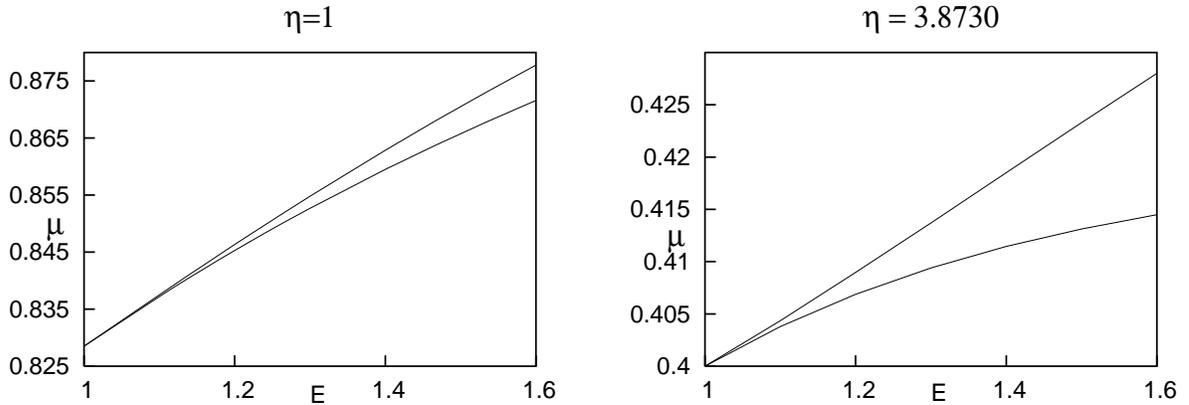}
\caption{These show higher order wedges of instability. Both are cases
of resonance between the two hydrodynamic modes but with
$\omega_1-\omega_2 =4$, rather than $2$. The left-hand panel shows the
instability wedge when $\eta = 1$, the right-hand panel shows the
corresponding wedge when $\eta = \sqrt{15}$.}\label{ho}
\end{figure}

In identifying these tongues, we have made repeated use of the
asymptotic formulas presented earlier. The numerical procedure also
picks up some further tongues, related to higher-order resonances, that are excluded by the procedure leading to the asymptotic formulas. These we
have mostly ignored on the ground that they are too weak and occupy too small
a region of the parameter space to be significant, but we show one such resonance tongue for two values of $\eta$ in Figure \ref{ho}
\placefigure{ho}

\section{Upper bound on fieldstrength}\label{sec:upperbound}

The results of \S\S \ref{kne2} and \ref{numerical} show that magnetoelliptic
instability occurs in vertically
unbounded systems whatever the fieldstrength, and that the growth rates are
relatively insensitive to the magnetic fieldstrength parameter $\eta$. The
loci of instability in the $(E, \mu)$ plane, however, do depend on $\eta$. As
$\eta\rightarrow\infty$, the resonant $\mu$, in all three cases,
scales as $\eta^{-1}$ and the magnetic tension parameter $m\rightarrow 1$. This
is a consequence of the resonant character of the instability. Since 
the mode frequency must be close to the rotational frequency, the Alfv\'en
frequency $m$ cannot be too large. As $B\rightarrow\infty$, $k_3$ must approach
zero.

In a system of finite vertical thickness $H$, $k_3$ cannot drop below $\chi/H$,
where $\chi$ is a factor of order unity. Therefore, if $v_A$ exceeds $\Omega
H/\chi$, we expect the magnetoelliptic instability to be suppressed. A precise
upper bound on $v_A$ follows from eqn. (A3) and application of the resonance
conditions; equations (\ref{pure_hydro_res}), (\ref{equ:reshm}), and
(\ref{equ:resmm}). 
In Cases 1 and 2, instability requires $m^2\le 1$. Case 3 requires
$m^2\le 3$. Therefore, magnetoelliptic instability requires $v_A\le\sqrt{3}
\Omega H/\chi$ (there is a correction of order $\epsilon$ due to the finite
bandwidth). The corresponding fieldstrength
 is comparable to the maximum value of the field at
which the magnetorotational instability can operate \citep{mri}.

\section{Discussion}\label{discussion}
We have explored the effect of a uniform, vertical magnetic field on
the stability of planar, incompressible flow with elliptical
streamlines in an unbounded medium, in the approximation of ideal
magnetohydrodynamics. In the absence of magnetic fields, flows with
elliptical streamlines having ellipticity parameter $\epsilon$ [see
equation (\ref{equ:definitions})] are known to be unstable to
perturbations with wavevectors that are transverse to the plane of the
flow (the ``elliptical instability"). Our first conclusion is that
this elliptical instability persists in the presence of a vertical
magnetic field: the latter decreases the maximum growthrate but fails
to suppress the instability, no matter how large the magnetic-field
parameter becomes. It can be compared with the conclusion of
\citet{rk94} that a toroidal magnetic field has a stabilizing
influence. Kerswell's analysis holds for small $\epsilon$ only and
shows that the growthrate decreases with magnetic field in that
limit. Our result, which holds for a vertical magnetic field, shows a
similar trend for small values of the magnetic-field parameter, but
this trend never results in complete stabilization of the elliptical
instability with increasing magnetic field.

A second conclusion is that there are further instabilities associated
with the presence of the magnetic field. One of these, for which the
eigenvector is a mixture of hydrodynamic and magnetic modes, occurs
for all values of the magnetic field parameter. Another, for which the
eigenvector is a combination of magnetic modes only, sets in for
values of the magnetic-field parameter exceeding a certain threshold
value ($\eta > \sqrt{3}$). 
In all three of these instabilities, for large magnetic fields, the
wave vector makes only a small angle with the plane of the
unperturbed flow, reflecting the familiar tendency for dynamics to become
nearly 2-dimensional in a strong, well ordered magnetic field. This is
reflected in Figure \ref{mixedmode}, which shows that as $\eta$ increases,
the unstable wedges are pushed to smaller $\mu$. Although the
unstable fraction of
the $(E,\mu)$ plane decreases with increasing $\eta$ (except for a very
slight maximum at $\eta\sim$ 2.18, reflecting the onset of instability
between magnetic modes), the separation between the unstable wedges also
decreases.  While
the nonlinear evolution of the unstable
system is beyond the scope of this work, the destabilization of a nearly
continuous swath of parameter space may have consequences for the 
interactions between unstable
modes.
In all three
cases, the maximum instability increment tends
to $\epsilon \pi/2$, i.e., the maximum growthrate of the unstable
modes tends, in dimensional units, to $\epsilon\Omega /4$. 

All three of these instabilities may be relevant in accretion-disk settings. 
In systems of finite thickness $H$, however, the instability is suppressed if
the Alfv\'en speed $v_A$ exceeds a critical value of order $\Omega H$.
Magnetorotational instabilities are quenched at approximately the same
fieldstrength \citep{mri}.
The growth rate of magnetoelliptical instabilities is smaller than
that of magnetorotational instabilities by a factor of order $\epsilon$, and
thus they are not necessarily the primary instability in magnetized disks.
They may well play a secondary
role by breaking up eddies or vortices generated by other
mechanisms. 

Magnetoelliptic instabilities may also occur in the inner parts of
barred galaxies, in which the gas flow is slightly elliptical and the
magnetic field, at least in the Milky Way, has a vertical component
\citep{mor}. In such settings, the instabilities could be a source of
turbulence, possibly affecting the mass supply to a central compact
object.

\acknowledgements

We are happy to acknowledge the referee for useful comments. Material
 support for this work was provided by NSF grants AST-0098701,
AST-0328821, PHY-0215581, and the Graduate School of the University of
Wisconsin, Madison.
\appendix

\pagebreak
\noindent {\Huge \bf Appendices} 

\noindent We carry out calculations leading to
the coefficients appearing in equation (\ref{beta1_eq}) in a series of
steps. Since $p(\lambda)= |M - \lambda I|$, with $M$ given by
equations (\ref{M_exp_2}) and (\ref{M0andM1}), we begin with the expression for $M$.

\section{The Expansion for M}
\subsection{Zero-order Problem and Solution for $M$}\label{subsec:zop}
The matrix $D\left(t, \epsilon , \mu\right)$ has only two entries
that depend on $\epsilon$: $D_{11}$ and $D_{21}$. If, therefore, we
write the Taylor expansion
\begin{equation}\label{D_expansion} D\left(t, \epsilon, \mu \right) = D_0 \left(t, \mu \right) + \epsilon D_\epsilon \left( t,\mu \right) + \cdots \end{equation}
then
\[ D_0 \left(t,\mu \right) = D\left(t, 0, \mu \right) \; \mbox{and}\; D_\epsilon \left(t, \mu \right) = D_\epsilon \left(t, 0, \mu \right), \]
where $D_\epsilon = \partial D/\partial \epsilon$.
For $D_0$ we find the constant matrix
\begin{equation}\label{D0_matrix}
D_0\left(t, \mu \right) = D_0\left(\mu \right)= \left(
\begin{array}{cccc}0&-2&im&0\\2\mu ^2&0&0&im\\im&0&0&0\\0&im&0&0
\end{array}\right).\end{equation} Its eigenvalues are
\begin{equation}\label{evals}
\sigma _1 = i\left(\mu + q\right),\sigma _2 = -i\left(\mu +
q\right),\sigma _3 = i\left(\mu - q\right),\sigma _4 = -i\left(\mu -
q\right),\end{equation} where $q = \sqrt{\mu ^2 + m^2} = \mu \sqrt{1 +
\eta ^2}.$ These are distinct and nonzero as long as $\mu \ne 0$ and
$\eta \ne 0$, which we assume to be the case.  The first two
correspond to ``hydrodynamic modes'' since they reduce, when $\eta=0$,
to the eigenvalues of the purely hydrodynamic case. The second two
refer to ``magnetic modes'' since they are zero in that limit. These
are all of stable type, corresponding to frequencies $\omega _k$, $k=
1, \ldots, 4$. Regarding the matrix $D_\epsilon$, one can easily work it out
from the expression (\ref{D_matrix}) above: all its entries vanish
except $\left(D_\epsilon\right)_{11}$ and $\left(D_\epsilon\right)_{21}.$ One finds
\begin{equation}\label{D11}\left(D_\epsilon\right)_{11} = i\left(1-\mu
 ^2\right)\left(e^{2it} - e^{-2it}\right),\end{equation} and

\begin{equation}\label{D21}\left(D_\epsilon\right)_{21} = \mu ^2 \left(1-\mu ^2 \right)
 \left(e^{2it} + e^{-21t} -2 \right) + 2\mu \nu.
 \end{equation}

\bigskip

tFrom the matrices $D_0$ and $D_\epsilon$ we can construct the matrices
$M_0\left(\mu \right)$ and $M_\epsilon\left(\mu\right)$ needed in the
formula (\ref{M0andM1}) for $M_1$. For $M_0$ we simply have $\exp {2\pi
D_0}$. For $M_\epsilon$ we proceed as follows. 
On the finite time-interval $[0,2\pi]$ we may write
\[ \Phi \left(t,\epsilon ,\mu \right) = \Phi _0\left(t,\mu\right) + \epsilon \Phi _1 \left(t, \mu \right) + \ldots, \; \Phi _1 \left(0, \mu \right)=0. \]
Substituting this in the
differential equation (\ref{c_eq}), expanding to first order in
$\epsilon$, using the variation of constants formula (cf. \citet{cl}) and setting $t=2\pi$, we get

\begin{equation}\label{floquet_matrix}
M\left(\epsilon , \mu \right) = M_0\left(\mu \right)\left(I + \epsilon \int _0^{2\pi} \Phi_0^{-1}\left(s, \mu \right)D_\epsilon\left(s,\mu \right) \Phi _0
\left(s, \mu \right)ds\right).\end{equation}
This expresses the Floquet matrix correctly to linear order in $\epsilon$, and this will turn out to be sufficient for our purpose. The formula above identifies $M_\epsilon\left(0,\mu\right)$:
\begin{equation}\label{M_epsilon}
M_\epsilon \left(0,\mu \right) = M_0 \left(\mu \right) \int _0^T \Phi
_0^{-1}\left(s, \mu \right)D_\epsilon\left(s,\mu \right) \Phi _0
\left(s, \mu \right)ds .\end{equation}
We next proceed to simplify this expression.

\subsection{A further transformation}\label{distinct}
The characteristic polynomial given in equation (\ref{charpol}) above 
is the same in any coordinate system, so we shall choose one to simplify the unperturbed Floquet matrix $M_0\left(\mu\right)$.

If $\mu \ne 0$ and $\eta \ne 0$ the eigenvalues $\{\sigma _k\}$ given
by equation (\ref{evals}) are all distinct, so the 
eigenvectors are linearly independent and the matrix $T\left(\mu\right)$ 
formed from their columns diagonalizes $D_0$:
\[\tilde{D}_0 =
\mbox{diag}\left(\sigma_1,\sigma_2,\sigma_3,\sigma_4\right),\]  
where the tilde indicates the transformed matrix: $\tilde{D}=T^{-1}DT$.
We shall need to know $T$ and $T^{-1}$ explicitly. 
It is a
straightforward matter to show that any eigenvector of $D_0$ must have
the structure (up to a constant multiple)
\[ \xi = \left(\begin{array}{c} \sigma \\-\left(\sigma ^2 +
m^2\right)/2\\im \\ -i m\left(\sigma ^2 + m^2 \right)/2\sigma
\end{array} \right).\]
Substituting the particular values of $\sigma$ given in equation
(\ref{evals}) gives the four columns of the matrix $T$, and from this
we can construct its inverse. One finds
\begin{equation}\label{Tmatrix}
T = \left(\begin{array}{cccc}\sigma _1&\sigma _2 & \sigma _3 &\sigma
_4\\ -i\mu \sigma_1 & i\mu \sigma_2 & -i\mu \sigma_3 & i\mu \sigma_4
\\ im & im & im& im \\m\mu & -m\mu & m\mu & -m\mu \end{array}\right),
\end{equation}
and 
\begin{equation}\label{Tinverse}
T^{-1} = \frac{1}{4mq} \left( \begin{array}{cccc} -im & m/\mu & \sigma
_3 & i\sigma _3/\mu \\ im & m/\mu & \sigma _3 & -i\sigma _3/\mu \\ im
& -m/\mu & -\sigma _1 & -i \sigma _1 /\mu\\ -im & -m/\mu & -\sigma _1 &
i\sigma _1/ \mu
\end{array}\right).\end{equation}
The matrices $T$ and $T^{-1}$ depend on $\mu$ and on $\eta$ through the parameters 
\begin{equation}\label{mqdef}m=\mu \eta \; \mbox{and} \; q=\sqrt{\mu ^2 + m^2} = \mu \sqrt{1+\eta ^2}.\end{equation}

\medskip
In place of equation (\ref{floquet_matrix}) we now obtain
\begin{equation}\label{Mtilde}
\tilde{M} = \tilde{M}_0\left(I + \epsilon \tilde{J}\right),
\end{equation}
where 
\begin{equation}\label{Jeq}
\tilde{J}\left(\mu \right) = \int _0^{2\pi}
\tilde{\Phi}^{-1}\left(t,\mu \right)
\tilde{D}_1\left(t,\mu \right) \tilde{\Phi} \left(t,\mu \right)
\,dt.\end{equation}
Because the eigenvalues $\{\sigma _k \}$ are distinct, the matrix
$\tilde{\Phi} = \exp \left\{ \tilde{D}_0 t \right\}$ takes the simple, 
diagonal form
\begin{equation}\label{fundmatdistinct}
\tilde{\Phi}\left(t\right) = \mbox{diag}\, \left(\exp \left(\sigma _1 t
\right),\exp \left(\sigma _2 t
\right),\exp \left(\sigma _3 t
\right),\exp \left(\sigma _4 t
\right) \right)
\end{equation}
The $ij$ entry of the matrix
$\tilde{D}_1$
is (since $D_\epsilon$ has only two nonzero entries)
\begin{equation}\label{D1ij}
\left(\tilde{D}_1\right)_{ij} = T_{1j}\left(\left( T^{-1}\right)_{i1}
\left(D_\epsilon\right)_{11} + \left(T^{-1}\right)_{i2} \left(D_\epsilon\right)_{21}\right).
\end{equation}
As a result, the $ij$ entry of the matrix $\tilde{J}$ providing the
leading-order perturbation of the Floquet matrix is
\begin{eqnarray}\label{Jij}
\tilde{J}_{ij} &=& T_{1j}\left(T^{-1}\right)_{i1}\int _0^{2\pi}
e^{\left(\sigma _j - \sigma _i
\right)t}\left(D_\epsilon\right)_{11}\left(t\right) \,dt \nonumber \\ &+&T_{1j}\left(T^{-1}\right)_{i2}\int _0^{2\pi}
e^{\left(\sigma _j - \sigma _i
\right)t}\left(D_\epsilon\right)_{21}\left(t\right) \,dt .
\end{eqnarray}
This enables us to find $\tilde{M}_\epsilon \left(0,\mu\right)$.

For the matrix $\tilde{M}_0\left(\mu\right)$ we have the expression 
\begin{equation}\label{tildeM0}
\tilde{M}_0\left(\mu\right)=\tilde{M}\left(0,\mu\right) = \mbox{diag} \left( \lambda _1 \left(\mu \right),  \ldots , \lambda _4\left(\mu  \right) \right)
\end{equation}
with $\lambda _k = 2\pi \sigma _k$. Recall that in order to construct $\tilde{M}_1$ we need also the derivative of this matrix with respect to $\mu$,
\begin{equation}\label{tildeM0prime}
\tilde{M}_0^\prime \left(\mu\right)=\tilde{M}_\mu \left(0,\mu\right) = \mbox{diag} \left( \lambda _1^\prime \left(\mu \right), \ldots , \lambda _4^\prime \left(\mu \right) \right).
\end{equation}
According
to equations (\ref{evals}) and (\ref{mqdef}), each eigenvalue $\sigma _k$ of $D_0$ is
linear in $\mu$. Therefore $\sigma _k ^\prime \left(\mu \right) =
\sigma _k \left(\mu \right)/\mu $. Since $\lambda _k = \exp \left(2\pi
\sigma _k \right)$, we have
\begin{equation}\label{lambda_prime}
\lambda _k ^\prime \left(\mu \right) = \lambda _k \left(\mu \right) 2
\pi \sigma _k \left(\mu \right) /\mu = \lambda _k 2\pi i \omega _k /\mu .\end{equation}

The formulas of this section allow one to determine the matrices
$\tilde{M}_0$ and $\tilde{M}_1$. To produce from these the
coefficients $p_j\left(\lambda\right)$ appearing in equations
(\ref{p_exp_2}) and (\ref{beta1_eq}) above, we need formulas for the
derivatives of a determinant. These are presented in Section \ref{detderivs} below and applied in the following section.

\subsection{The Expansion for $p\left(\lambda,\epsilon\right)$}\label{subsec:charpoly}

We can now find the required expansion for $p\left(\lambda , \epsilon
\right)$ by identifying the matrix $A$ of Section \ref{detderivs} below with $\tilde{M} -\lambda I$ and
the coefficients $q_k$ with the coefficients $p_k \left(\lambda
\right)$ of equation (\ref{p_exp_2}). We obtain these coefficients by
writing $a_k = \lambda _k - \lambda,$ $A_{kl} ^\prime\left(0\right) =
\left(\tilde{M}_1\right) _{kl}$ and $A_{kl} ^{\prime
\prime}\left(0\right) = 2\left(\tilde{M}_2\right) _{kl}$. This gives
for $p_1$ the following expression (with $n=4$):

\begin{eqnarray}\label{p1}
 p_1 \left(\lambda\right) &=&
\left(\tilde{M}_1\right)_{11}\left(\lambda _2- \lambda \right)
\left(\lambda _3 - \lambda \right)  \left(\lambda _4 - \lambda
\right) \\&+& \left(\tilde{M}_1\right)_{22}\left(\lambda _1 - \lambda
\right) \left(\lambda _3 - \lambda \right)  \left(\lambda _4 -
\lambda \right) + {\nonumber}\\&+&
\left(\tilde{M}_1\right)_{33}\left(\lambda _1- \lambda \right)
\left(\lambda _2 - \lambda \right)\left(\lambda _4 -
\lambda \right){\nonumber}\\&+&
\left(\tilde{M}_1\right)_{44}\left(\lambda _1- \lambda \right)
\left(\lambda _2 - \lambda \right)\left(\lambda _3 -
\lambda \right){\nonumber}
,\end{eqnarray} and there is a similar,
lengthier expression for $p_2$ obtained by making the corresponding substitutions in  equation (\ref{det_der_2}) below. 

The development thus far has required
no assumptions regarding the multipliers $\left\{\lambda _k \right\}$.
We now suppose that $\lambda _1 = \lambda _2$. It is then clear from
the expression above that $p_1\left(\lambda _1 \right) =0,$ as
asserted in Section \ref{char_poly}. The coefficients appearing in
equation (\ref{beta1_eq}) are now easily found to be
\begin{equation}\label{p0pp_eq}
p_0^{\prime \prime }\left(\lambda _1\right) = 2\left(\lambda _3 -
\lambda _1\right) \left( \lambda _4 - \lambda _1 \right),
\end{equation}
\begin{equation}\label{p1p_eq}
p_1^\prime \left(\lambda _1\right) =
-\left\{\left(\tilde{M}_1\right)_{11} +
\left(\tilde{M}_1\right)_{22}\right\} \left(\lambda _3 - \lambda
_1\right)\left( \lambda _4 - \lambda _1 \right) \end{equation}
and
\begin{equation}\label{p2_eq}
p_2 \left(\lambda _1\right) = \left| \begin{array}{cc}
\left(\tilde{M}_1\right)_{11}&\left(\tilde{M}_1\right)_{12}\\\left(\tilde{M}_1\right)_{21}&\left(\tilde{M}_1\right)_{22}
\end{array} \right| \left(\lambda _3 - \lambda
_1\right) \left( \lambda _4 - \lambda _1
\right). \end{equation} Equation (\ref{beta1_eq}) therefore takes the
form
\begin{equation}\label{beta1_eq_2}
\beta _1 ^2 + \left\{\left(\tilde{M}_1\right)_{11} -
\left(\tilde{M}_1\right)_{22}\right\} \beta _1 + \left|
\begin{array}{cc}
\left(\tilde{M}_1\right)_{11}&\left(\tilde{M}_1\right)_{12}\\\left(\tilde{M}_1\right)_{21}&\left(\tilde{M}_1\right)_{22}
\end{array} \right| = 0 .
\end{equation}

We note that the calculation of the perturbation of $\lambda _1$ to first order in $\epsilon$, which
requires expanding
$p$ to second order, requires the expansion of the
Floquet matrix $\tilde{M}\left(\epsilon\right) = \tilde{M}_0 +
\tilde{M}_1 \epsilon + \cdots $ only to first order.

We have assumed that the coincident roots are the first two, $\lambda
_1 = \lambda _2$. If instead we should have $\lambda _k = \lambda _l$,
equation (\ref{beta1_eq_2}) is modified by the replacement $(1,2) \to
(k,l)$.

Equation (\ref{beta1_eq_2}), together with equations (\ref{Mtilde}),
(\ref{tildeM0}), (\ref{tildeM0prime}) and (\ref{lambda_prime}) leads
to equation (\ref{beta1_eq_3}) of the text. What remains is to
evaluate the integrals defining $\tilde{J}$, and we now turn to this.

\subsection{Calculating the Elements of $\tilde{M}_1$}\label{subsec:elements}
By equations (\ref{M0andM1}) and (\ref{M_epsilon}) above (see also
equation \ref{Jeq}), the matrix $\tilde{M}_1$ is given by the formula
\begin{equation}\label{M1tilde_def}
\tilde{M}_1 = \tilde{M}_0 \left(\mu \right)\tilde{J} + \nu \tilde{M}_\mu \left(0,\mu \right)
\end{equation}
where the entries of $\tilde{J}$ are given by equation
(\ref{Jij}). From equations (\ref{D11}) and (\ref{D21})
it is a straightforward matter to carry out the integrations. We'll use for $T$ the matrix given above in equation (\ref{Tmatrix}). Since for this matrix $T_{1j} = \sigma _j= i \omega _j$, the formula for the entries of $\tilde{J}$ becomes
\begin{equation}\label{newJtilde}
\tilde{J}_{ij} = \sigma _j \left\{ \left(T^{-1}\right) _{i1} \int _0
^{2\pi} e^{\left(\sigma _j - \sigma _i \right) t }
\left(D_\epsilon\right)_{11}\left(t\right) \,dt + \left(T^{-1}\right)_{i2}
\int _0^{2\pi} e^{\left(\sigma _j - \sigma _i \right) t
}\left(D_\epsilon\right)_{21} \,dt\right\} .
\end{equation}

\medskip

For the diagonal entries the exponential factors in the integrand reduce to unity and one finds
\begin{equation}\label{diagonal_eq}
\tilde{J}_{jj} = -4\pi \mu^2 \left(1-\mu ^2 \right) \sigma _j \left(T^{-1} \right) _{j2},
\; j=1,2,3,4.\end{equation}

\medskip
For the off-diagonal entries, the formulas may be found generally, but
we need them only in the resonant cases where, for some pair of
indices $(i,j)$, $\sigma _i - \sigma _j = ki$ for a non-zero integer
$k$.\footnote{Recall that $\sigma _i \ne \sigma _j$ for any pair
$i,j$ so $k=0$ is excluded.} It is clear from the formulas (\ref{D11}) (\ref{D21}) and
(\ref{newJtilde}) that only resonances with $k=\pm 2$ contribute
off-diagonal terms to leading order in $\epsilon$ since for any other
choice of $k$ the integrals vanish: for $k \ne 2$ the matrix
$\tilde{J}$ is diagonal.

\section{Determinantal Derivatives}\label{detderivs}

Consider an $n
\times n$ matrix $A\left(\epsilon \right)$ having the properties that
it is a smooth function of $\epsilon$ and is diagonal at $\epsilon =
0$: $A\left(0\right) = \mbox{diag}\,\left(a_1,a_2, \ldots ,
a_n\right)$.

We need the coefficients in the Taylor expansion of
det$A\left(\epsilon\right) \equiv q\left(\epsilon\right)$:
\begin{equation}\label{q_exp}
q\left(\epsilon \right) = q_0 + q_1 \epsilon + q_2 \epsilon ^2 +
\cdots
\end{equation}
where
\begin{equation}\label{detA}
q_0= \left|A\left(0\right)\right|, \; q_1 = \frac{d
\left|A\left(\epsilon\right)\right|}{d\epsilon}|_{\epsilon =0}, \; q_2 =
\frac{1}{2} \frac{d^2
\left|A\left(\epsilon\right)\right|}{d \epsilon ^2}|_{\epsilon =0}, \;
\cdots\end{equation}
Straightforward applications of the formula for the derivative of a
determinant show that
\begin{eqnarray}\label{det_der_1}
q_1 &=& A_{11}^\prime   a_2 a_3 \cdots a_n +
A_{22}^\prime   a_1 a_3 \cdots a_n + \cdots +
A_{nn}^\prime   a_1 a_2 \cdots a_{n-1},\\
q_2 &=& \frac{1}{2}\left[ A_{11}^{\prime \prime}  a_2 a_3 \cdots a_n + 
 A_{22}^{\prime \prime}  a_2 a_3 \cdots a_n + \cdots +
A_{nn}^{\prime \prime}  a_1 a_2 \cdots a_{n-1}\right] \nonumber \\ &+& \left|
\begin{array}{cc}A_{11}^\prime &A_{12}^\prime
\\A_{21}^\prime & A_{22}^\prime
\end{array}\right| a_3 a_4 \cdots a_n +\left|
\begin{array}{cc}A_{11}^\prime &A_{13}^\prime
\\A_{31}^\prime & A_{33}^\prime
\end{array}\right|a_2 a_4 \cdots a_n  \\ &+&  \cdots +\left|
\begin{array}{cc}A_{(n-1)(n-1)}^\prime &A_{(n-1)n}^\prime
\\A_{n(n-1)}^\prime & A_{nn}^\prime
\end{array}\right|a_1 a_2 \cdots a_{n-2} \label{det_der_2},
\end{eqnarray}
where the terms involving two-by-two determinants represent the sum over $k < l$ of the
product of the $\left\{a_j\right\}$, $a_k$ and $a_l$ omitted, with the
determinant 
\[ \left| \begin{array}{cc}  A_{kk}^\prime   & A_{kl}^\prime
 \\A_{lk}^\prime   & A_{ll}^\prime
 \end{array}\right|\]
and all derivatives are evaluated at the origin.

\end{document}